\documentclass[twocolumn,showpacs,prb]{revtex4-1}

\usepackage{textcomp}
\usepackage{amsmath}
\usepackage{amssymb}
\usepackage{graphicx}
\usepackage{color}

\begin{document}

\title{Nonlinear response in a non-centrosymmetric topological insulator}

\author{Zhou Li$^{1,2}$}

\email{zli5@ualberta.ca}

\author{Franco Nori$^{1,3}$}

\affiliation{$^{1}$ Theoretical Quantum Physics Laboratory, RIKEN Cluster for Pioneering Research, Wako-shi, Saitama 351-0198, Japan \\
$^{2}$ Department of Electrical Engineering, Purdue University, West Lafayette,
Indiana 47906, United States\\
 $^{3}$ Department of Physics, University of Michigan, AnnArbor,
Michigan 48109-1040, United States}
\begin{abstract}
Nonlinear phenomena are inherent in most systems in nature. Second
or higher-order harmonic generations, three-wave and four-wave mixing
are typical phenomena in nonlinear optics. To obtain a nonzero signal
for second-harmonic generation in the long-wavelength limit ($q\rightarrow0$),
the breaking of inversion symmetry is required. In topological materials,
a hexagonal warping term which breaks the rotation symmetry of the
Fermi surface is observed by angular-resolved photo-emission spectroscopy
(ARPES). If a gap opens (e.g., by doping with magnetic impurities)
the inversion symmetry will be broken. Here we use a nonlinear response theory based on a generalized Kubo formula to explain the frequency up-conversion in topological materials. 
\end{abstract}

\date{\today}

\maketitle

\section{introduction}

The nonlinear response to an external driving electromagnetic field $\mathbf{E}$
or $\mathbf{B}$ can be characterized by a conductivity tensor $\tilde{\sigma}$ which
is not a constant, but depends on the magnitude of $\mathbf{E}$ or $\mathbf{B}$. Nonlinearity
is often found to be important in optical devices, especially in the
recent discovery of high-efficiency solar energy harvesting
in non-centrosymmetric crystal structures such as perovskite oxides \cite{Grinberg,Nie,Shi,Quile,Cook}. In 3D topological insulator (TI) and ferroelectric materials, Dirac
cones \cite{Hasan,Qi1,Moore,Hsieh1,Chen} obeying spin-momentum
locking \cite{Hsieh2,Nori} with in-plane spin component perpendicular
to the momentum $\mathbf{k}$ were verified by spin-sensitive angular-resolved
photo emission spectroscopy (ARPES). The quasiparticles (helical Dirac
fermions) observed in topological materials possess an important feature:
the Fermi contours are circular for small values of the chemical potential
$\mu$, and acquire a snowflake \cite{Chen} shape as $\mu$ increases.
Analyzing the experiment results, Fu \cite{Fu} assigned a hexagonal
warping term in the Hamiltonian of such quasiparticles. This term
has a strong signature in the optical conductivity, spin texture,
Hall conductivity and circular dichroism of topological insulators \cite{Li1,Xiao3,Li2}.

The optical conductivity was predicted to show \cite{Li1} a large
near-linear increase with photon energy above the inter-band threshold
as compared to the usual flat background \cite{Carbotte1,Li08,Orlita}
inter-band optical conductivity in graphene. The spin texture \cite{Xu1}
(specifically, out of plane spin $S_{z}$) shows a mixture of up-and-down
directions; in contrast to the normal all-up or all-down hedgehog type \cite{Xu2} distribution for massive Dirac fermions (see, e.g. Fig.~5 of Ref.~[\onlinecite{Li2}]). It is also possible to introduce a gap in the topological surface
quasiparticles (massive Dirac fermions) by magnetic doping \cite{Chen1,Tokura}
in $\textrm{B\ensuremath{\textrm{i}_{2}}S\ensuremath{\textrm{e}_{3}}}$
\cite{Chen1} and recently in $\textrm{Cr}_{x}(\textrm{Bi}_{1-y}\textrm{Sb}_{y})_{2\text{\textminus}x}\textrm{Te}_{3}$ \cite{Tokura,Yasuda}. Considerable particle-hole asymmetry of the
surface Dirac cone of a 3D TI usually displays, which can be modeled
with a small sub-dominant Schr\"odinger quadratic-in-momentum term in addition to the dominant Dirac Hamiltonian.
While perhaps small, the Schr\"odinger term has been shown to provide
important modifications \cite{Li3} in the chiral nonlinear magneto
optical conductivity (MOC) which is related to the absorption of left
and right circularly polarised light of 3D TI. This is to be compared
with what is found in graphene \cite{Carbotte4,Xiao1,Yao} or the
related single layer silicene \cite{Tabert}.

In this work we focus on the nonlinear optical
conductivity induced by an electric field $\mathbf{E}$ in contrast to the
nonlinear MOC which is induced by a magnetic field $\mathbf{B}$. We consider three-wave mixing \cite{Shen} (e.g., second-harmonic generation)
from non-centrosymmetric topological materials. Second-harmonic generation (SHG)
was first demonstrated by projecting a laser beam
through crystalline quartz \cite{Franken}. Later on this effect was found in other materials (e.g., silicon surfaces) \cite{Heinz} with broken inversion symmetry. Theoretically, SHG was predicted
to be nonzero in semiconductors \cite{Sipe}, and more recently in
single-layer graphene \cite{Mikhailov,Daria} with oblique incidence
of radiation on the 2D electron layer. For oblique incidence, the
incident radiation has a nonzero wave vector component $\mathbf{q}$
parallel to the plane of the 2D layer. In the long-wavelength limit
($\mathbf{q}\rightarrow0$, normal incidence), the SHG vanishes because
graphene is a centrosymmetric material. However, higher-order harmonics
(e.g., third-harmonic generation) could be nonzero in graphene \cite{Wright}
or generally Dirac Fermions system \cite{Nagaosa}. The nonlinear coupling
of three monochromatic waves, thus called three-wave mixing, has been successfully used to generate optical frequency up-conversion
or down-conversion. Nonlinear optical analogs, including SHG, have also been studied recently in various contexts, including Josephson plasma waves \cite{Savel} and cavity quantum electrodynamics \cite{Kockum,Kockum1,Stassi,Gu}.

In the following paragraphs we present a Green's function formalism for calculating the nonlinear
conductivity in section II. We use a two-band hexagonal warping model which can be found in surface
states of 3D TI and ferroelectric materials. The inversion symmetry of the Fermi surface is broken by a magnetic doping in the hexagonal warping model. In section III we present the linear optical conductivity from the Green's function formalism. In section IV we present our numerical results of the nonlinear and linear conductivity for different sets of parameters (e.g., chemical potential, gap parameter, temperature, etc.). In section V we summarize our results with a conclusion.

\section{Nonlinear optical conductivity}

The linear conductivity $\tilde{\sigma}_{xx}(\omega) $ is related to the current $ J_x(\omega) = \tilde{\sigma}_{xx}(\omega)E_x(\omega)$, 
while the nonlinear conductivity $ \tilde{\sigma}_{xxx}(\omega,\omega) $ is related to the current $ J_x(2\omega) = \tilde{\sigma}_{xxx}(\omega,\omega)E^2_x(\omega)$.
In general, the nonlinear conductivity is a tensor $\tilde{\sigma}_{\alpha\beta\gamma}$. However, here for
simplicity we only consider the $xxx$ component of the tensor; the
other components of the conductivity tensor can be obtained in a similar
way. The nonlinear conductivity has been well studied in earlier references; for example, in Ref.~[\onlinecite{Bloemb}] the Eq.~(2-48) defines the nonlinear conductivity as a product of momentum matrix elements and then in Eq.~(2-49) the momentum matrix elements were connected to velocity matrix elements. 

For the linear conductivity it has been shown in chapter 8 of the book [\onlinecite{Economou}] that the Eq.~(8.53) uses a trace of momentum operators and Green's functions and then in Eq.~(8.55) this was connected to the product of velocity matrix elements. The velocity matrix element is connected to the position matrix element and the shift vector. \cite{ZhouLi} For the nonlinear conductivity, instead of using velocity matrix elements \cite{Bloemb} directly, we define the nonlinear conductivity as a trace of velocity operators and Green's functions \cite{Economou}; the imaginary frequency in each Green's function is set by using a triangle Feynman diagram.

\begin{widetext}
\begin{equation}
\tilde{\sigma}_{xxx}(\omega,\omega)=\frac{e^{3}}{\omega^{2}}\frac{i}{4\pi^{2}}\int_{0}^{2\pi}d\theta\int_{0}^{k_{\mathrm{cut}}}kdkT\sum_{l}\mathrm{Tr}\langle v_{x}\widehat{G}(\mathbf{k,}i\omega_{l})v_{x}\widehat{G}(\mathbf{k,}i\omega_{l}+i\omega_{n})v_{x}\widehat{G}(\mathbf{k,}i\omega_{l} - i\omega_{n})\rangle_{i\omega_{n}\rightarrow\omega+i\delta} \,,
\end{equation}
here $v_{x}$ is the velocity operator and $\widehat{G}(\mathbf{k,}i\omega_{l})$
is the matrix Green's function, $e$ is the charge of the electron,
$k$ the absolute value of the momentum $\mathbf{k}$ with direction
$\theta$ and cutoff $k_{\mathrm{cut}}$, $T$ is the temperature with $\omega_{n}=2n\pi T$, $\omega_{l}=(2l+1)\pi T$ the Boson and Fermion Matsubara frequencies, $n$ and $l$ are integers and $\mathrm{Tr}$ is a trace. To obtain the nonlinear conductivity, which is a real frequency quantity, we needed to make an analytic continuation from imaginary $i\omega_{n}$ to real $\omega$ and $\delta$ is infinitesimal. This is valid for the long wavelength limit $q\rightarrow0$; 

Consider a two-band model as an example, the velocity operators and matrix Green's functions are 2 $\times$ 2 matrices, and can be
expanded onto the basis of Pauli matrices $\mathbf{\sigma}=(\sigma_{x},\sigma_{y},\sigma_{z})$,
as $v_{x}=a_{0}+\mathbf{a}\cdot\mathbf{\sigma}$, and $\widehat{G}(\mathbf{k},i\omega_{n})=g_{0}+\mathbf{g}\cdot\mathbf{\sigma}$. We can use the algebra $(\mathbf{a}\cdot\mathbf{\sigma})(\mathbf{g}\cdot\mathbf{\sigma})=(\mathbf{a}\cdot\mathbf{g)}I_{2}+i(\mathbf{a}\times\mathbf{g)\cdot\mathbf{\sigma}}$
to evaluate the trace and the complicated results will be contained
in the function $F(k,\theta)$ to be integrated further in momentum
space $(k,\theta)$. We can also perform the sum over the internal
Fermion Matsubara frequencies $\omega_{l}$ and the result is Fermi-Dirac
distribution function defined as $f(x)=1/[\exp(x/T-\mu/T)+1]$. After tedious but straightforward algebra (details in the appendix), we finally obtained both the
inter-band and intra-band nonlinear optical conductivity, the intra-band optical conductivity contributes to the frequency region of $\omega\approx0$ and was given in the appendix, in the equations below we present the results of the inter-band optical conductivity ($\tilde{\sigma}^{\textrm{inter}}_{xxx}=\tilde{\sigma}^{\textrm{inter}}_{xxx}(\omega,\omega)$),
\begin{eqnarray}
\tilde{\sigma}^{\textrm{inter}}_{xxx} = \frac{ie^{3}}{\hbar^3 \omega^{2}\pi^{2}}\int kdkd\theta F(k,\theta)\frac{[f(E)-f(-E)]}{E}\Big[\frac{1}{\hbar \omega+i\delta+2E}-\frac{1}{\hbar \omega+i\delta-2E}+\frac{2}{\hbar \omega+i\delta-E}-\frac{2}{\hbar \omega+i\delta+E}\Big]
\end{eqnarray}
\end{widetext}
Here $E$ is the quasiparticle energy which depends on the momentum
$(k_x,k_y)=(k\mathrm{cos}(\theta),k\mathrm{sin}(\theta))$. Take a two-band hexagonal warping model as an example,
the Hamiltonian is given by, 
\begin{equation}
H_{0}=v_{k}(k_{x}\sigma_{y}-k_{y}\sigma_{x})+\frac{\lambda}{2}(k_{+}^{3}+k_{-}^{3})\sigma_{z}+M\sigma_{z}\,,
\end{equation}
this model has been used to describe the surface states band structure
near the $\Gamma$ point in the surface Brillouin zone of a 3D TI
and also recently in ferroelectric materials. The Dirac fermion velocity
to second order is $v_{k}=\hbar v_{\mathrm{F}}(1+\alpha k^{2})$, with $v_{F}$
the usual Fermi velocity and $\hbar v_{\mathrm{F}}$ measured to be 2.55 eV$\cdot\mathrm{\mathring{A}}$
and $\alpha$ is a constant which is fit along with $m$ to the measured
band structure in Ref.~[\onlinecite{Fu}]. Here $m$ appears in the quadratic term $\hbar^2 k^2/(2m)$ which, for simplicity, is dropped in the Hamiltonian $H_{0}$. The inclusion of the quadratic term provides particle-hole asymmetry; however the wave function is not changed \cite{Li2}, thus the Berry curvature and Berry connection (defined from the wave function) are not modified by this quadratic term. For simplicity, the quadratic correction to the velocity $\alpha$ is also discarded. The magnitude of the hexagonal warping parameter is $\lambda=$200 eV$\cdot\mathrm{\mathring{A}}^{3}$, estimated from the measured Fermi velocity. The same value was used in Ref.~[\onlinecite{Fu}]. The $\sigma_{x}$, $\sigma_{y}$,
$\sigma_{z}$ are Pauli matrices here referring to spin, while in graphene
these would relate instead to pseudospin. Finally $k_{\pm}=k_{x}\pm ik_{y}$, with the $k_{x}$, $k_{y}$ momentum along the $x$ and $y$ axis, respectively.
$M$ is the strength of the gap opened when the topological thin film
is in proximity to magnetic impurities.

\begin{figure}[tp]
\begin{centering}
\includegraphics[width=2.8in,height=3.5in]{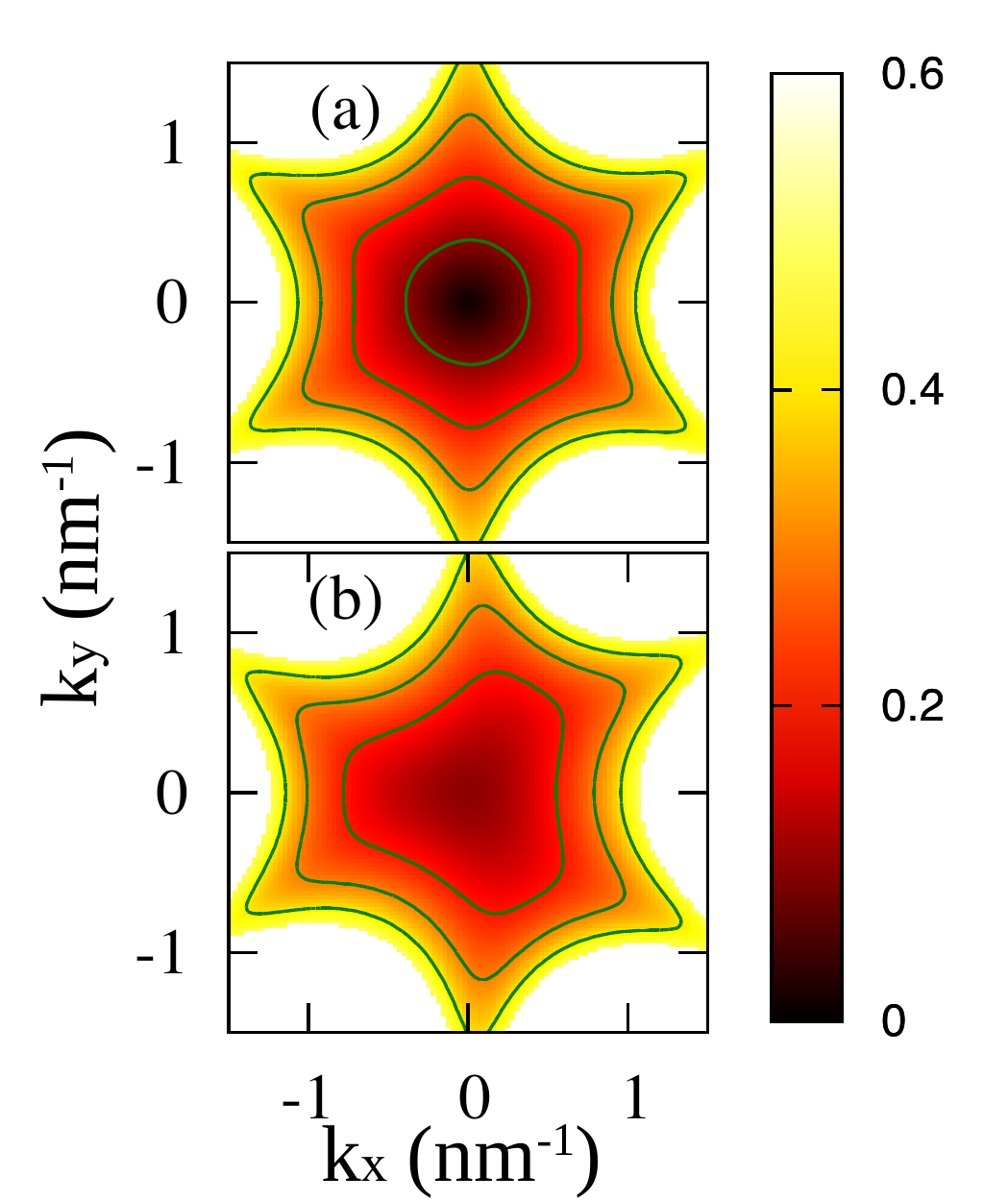} 
\end{centering}
\caption{(Color online) Constant-energy contours for the dispersion curves
used to describe the surface states in 3D TI. In (a) the
gap is $2M$=20 meV, while in (b) the gap is $2M$=200 meV.
The chemical potential $\mu$ can be changed by doping the 3D TI.
The four contours shown in green correspond to different
chemical potentials $\mu=0.1$ eV (in (b) this contour
disappears), $\mu=0.2$ eV, $\mu=0.3$ eV and $\mu=0.4$ eV. The $k_{x}$
and $k_{y}$ axes are in units of 0.1 $\mathrm{\mathring{A}}^{-1}$. In
(a), for the contour $\mu=0.1$ eV, one can see the Fermi
surface deviate slightly from a perfect circle; and for higher
chemical potential, the Fermi surface is a snowflake shape. In (b) the Fermi surface becomes significantly distorted. The
hexagonal warping parameter here is $\lambda=$0.2 $\textrm{eV}\cdot(\textrm{nm})^{3}$.}
\label{fig1} 
\end{figure}

The quasiparticle energy dispersion relation is given by $E=\sqrt{v_{k}^{2}k^{2}+[\lambda k^{3}\cos(3\theta)+M]^{2}}$,
and the function $F(k,\theta)$ is given by 
\begin{eqnarray}
F(k,\theta) & = & \frac{[k_{x}v_{k}^{2}+3\lambda(k_{x}^{2}-k_{y}^{2})(M+\lambda k_{x}(k_{x}^{2}-3k_{y}^{2}))]}{E^{3}}\notag\\
 & \times & v_{k}^{2}[v_{k}^{2}k_{y}^{2}+\lambda^{2}(4k_{x}^{6}+9k_{x}^{4}k_{y}^{2}-18k_{x}^{2}k_{y}^{4}+9k_{y}^{6})\notag\\
 &  & -4\lambda Mk_{x}^{3}+M^{2}]\,.
\end{eqnarray}
Note that if $M=0$ or $\lambda=0$ the integration $\int_{0}^{2\pi}d\theta\int_{0}^{k_{\mathrm{cut}}}kdk$
will be zero because the integrand is an odd function of $k_{x}$.
So only when both $M\neq0$ and $\lambda\neq0$ we obtain a non-vanishing
second-harmonic generation nonlinear conductivity in the long-wavelength
limit $q\rightarrow0$.

\section{Linear optical conductivity}

It is well known \cite{Li2} that the linear optical conductivity is obtained from the standard Kubo formula in terms of the matrix Green's function and velocity operators, the
longitudinal conductivity is given by
\begin{eqnarray}
&&\tilde{\sigma} _{xx}(\omega )=\frac{e^{2}}{i\omega }\frac{1}{4\pi ^{2}}
\int_{0}^{k_{cut}}kdkd\theta \times   \notag \\
&&T\sum_{l}\mathrm{Tr}\langle v_{x}\widehat{G}(\mathbf{k,}i\omega _{l})v_{x}\widehat{G}
(\mathbf{k,}i\omega _{n}+i\omega _{l})\rangle _{i\omega _{n}\rightarrow \omega
+i\delta }
\end{eqnarray}
which works out to be
\begin{eqnarray}
&&\tilde{\sigma}^{\textrm{inter}} _{xx}(\omega )=\frac{ie^{2}}{4\pi ^{2}\hbar^2 \omega }
\int_{0}^{k_{cut}}kdkd\theta H(k,\theta ) \notag \\
&&\frac{f(E)-f(-E)}{E}\Big[\frac{1}{\hbar \omega+i\delta +2E} -\frac{1}{\hbar \omega+i\delta-2E}\Big]
\end{eqnarray}
where the function $H(k,\theta )$ is given by
\begin{eqnarray}
&&H(k,\theta )=\frac{v_{k}^{2}}{E}[9\lambda ^{2}k^{6}\cos ^{2}(2\theta)+(M+\lambda k^{3}\cos (3\theta ))^{2}+  \notag \\
&&v_{k}^{2}k^{2}\sin ^{2}\theta-6\lambda k^{3}\cos (2\theta )\cos \theta(M+\lambda k^{3}\cos (3\theta ))]
\end{eqnarray}
It is interesting to check the units of $H(k,\theta)$ and $F(k,\theta)$. We find that $F(k,\theta)\times k$ and $H(k,\theta)$ have the same unit as $v_{k}^{2}E$. So $\tilde{\sigma} _{xxx}$ has the same unit as $\tilde{\sigma} _{xx}\times (e/\hbar\omega k) $. Then the product of the nonlinear conductivity $\tilde{\sigma} _{xxx}$ and external electric field, $\tilde{\sigma} _{xxx}\times E_x$ has the same unit as $\tilde{\sigma} _{xx}$, as expected.

\section{Numerical Results}
To evaluate the nonlinear optical conductivity, we need to perform an integration in momentum space which is restricted by the Fermi-Dirac
distribution function $f(x)$. At zero temperature, the restricted area is the Fermi surface shown in Fig.~1. In (a), for a small chemical potential $\mu=0.1$ eV, the Fermi surface is very close
to but not a perfect circle because of the small gap $2M=20$ meV. At larger chemical potential $\mu=0.4$ eV, the Fermi surface deviates a bit from a snowflake shape. In (b) a much larger gap $2M=200$ meV is used and the Fermi surface is significantly distorted. The inversion symmetry is broken in both (a) and (b).

\begin{figure}[t!]
\begin{centering}
\includegraphics[width=3.5in,height=3.in]{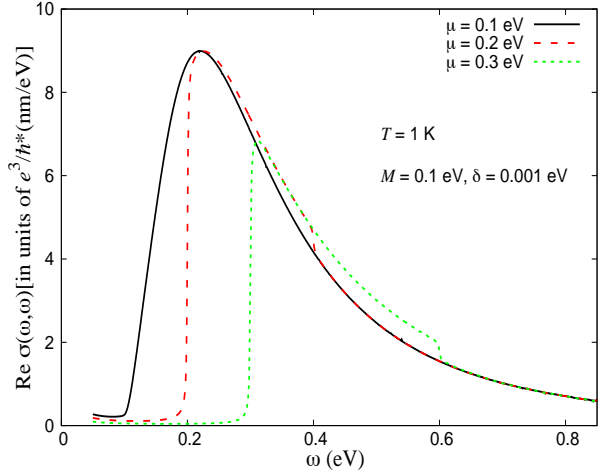} 
\end{centering}

\caption{(Color online) Real part of the nonlinear optical conductivity $\tilde{\sigma}_{xxx} (\omega,\omega)$ versus frequency $\omega$ in eV. Three different chemical potentials are used, with black solid for $\mu=0.1$ eV, red dashed for $\mu=0.2$ eV and green short dashed for $\mu=0.3$ eV. The impurity scattering self-energy $\delta=0.001$ eV and the temperature  $T=1$ K. }
\label{fig2} 
\end{figure}

\begin{figure}[t!]
\begin{centering}
\includegraphics[width=3.5in,height=3.in]{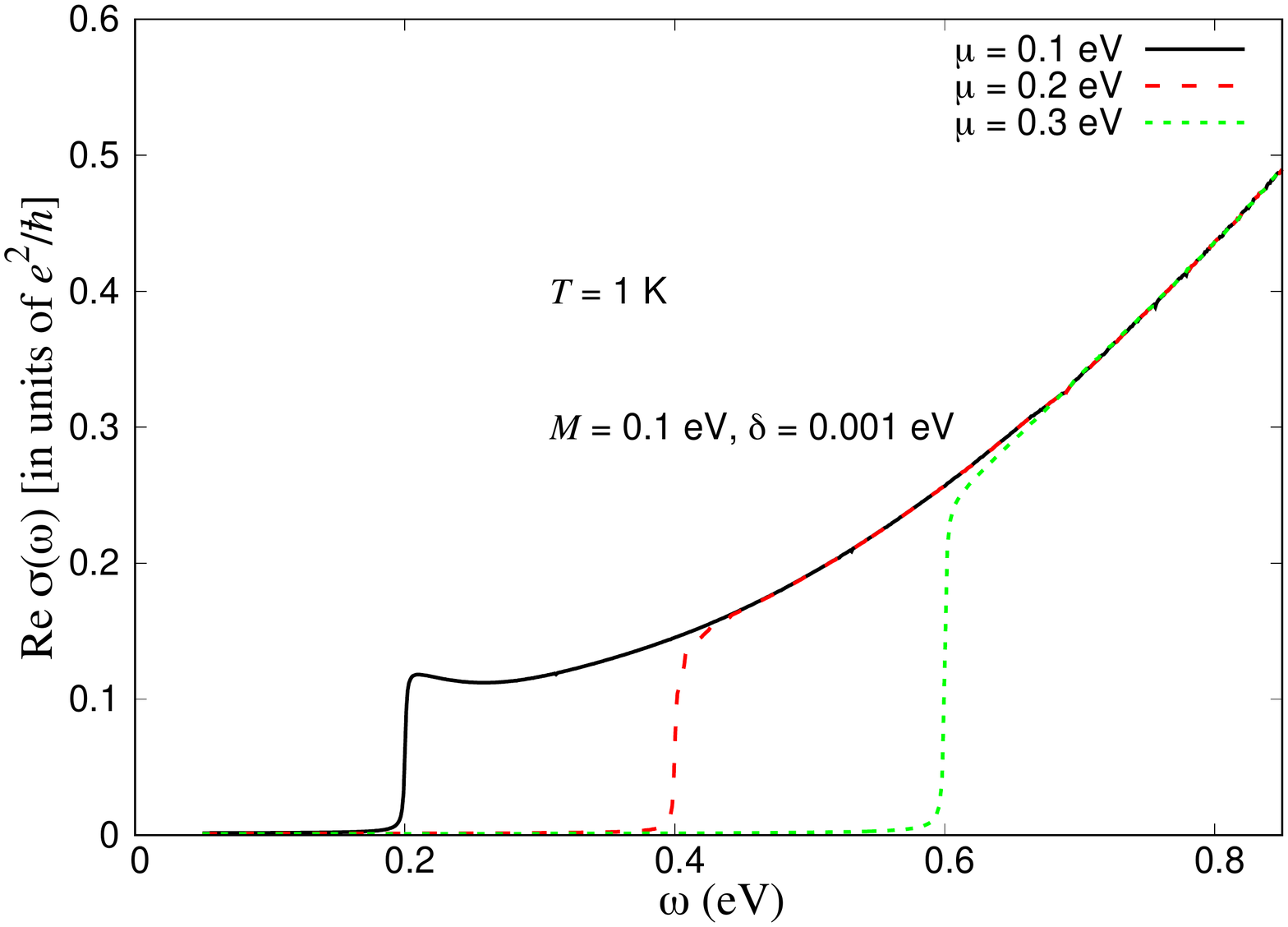} 
\end{centering}

\caption{(Color online) Real part of the linear optical conductivity $\tilde{\sigma}_{xx} (\omega)$ versus frequency $\omega$ in eV. Three different chemical potentials are used, with black solid for $\mu=0.1$ eV, red dashed for $\mu=0.2$ eV and green short dashed for $\mu=0.3$ eV. The impurity scattering self-energy $\delta=0.001$ eV and the temperature  $T=1$ K. }
\label{fig3} 
\end{figure}

\begin{figure}[t!]
\begin{centering}
\includegraphics[width=3.5in,height=3.in]{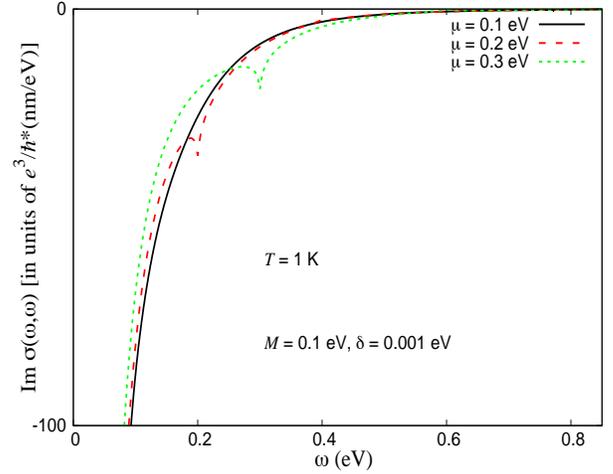} 
\end{centering}

\caption{(Color online) Imaginary part of the nonlinear optical conductivity $\tilde{\sigma}_{xxx} (\omega,\omega)$ versus frequency $\omega$ in eV. Three different chemical potentials are considered, with black solid for $\mu=0.1$ eV, red dashed for $\mu=0.2$ eV and green short dashed for $\mu=0.3$ eV. The impurity scattering self energy $\delta=0.001$ eV and the temperature  $T=1$ K. The sharp drops at  $\omega=$ 0.2eV and 0.3eV correspond to the sharp jump in the real part of the conductivity shown in Fig.~2. }
\label{fig4} 
\end{figure}

\begin{figure}[t!]
\begin{centering}
\includegraphics[width=3.5in,height=3.in]{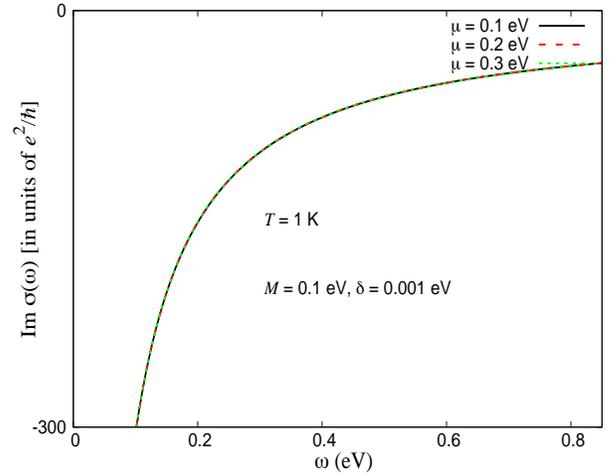} 
\end{centering}

\caption{(Color online) Imaginary part of the linear optical conductivity $\tilde{\sigma}_{xx} (\omega)$ versus frequency $\omega$ in eV. Three different chemical potentials are used, with black solid for $\mu=0.1$ eV, red dashed for $\mu=0.2$ eV and green short dashed for $\mu=0.3$ eV. The impurity scattering self energy $\delta=0.001$ eV and the temperature  $T=1$ K. }
\label{fig5} 
\end{figure}

In Fig.~2 and Fig.~3 we plot the numerical results of the real part of the inter-band nonlinear optical conductivity $\tilde{\sigma}^{\textrm{inter}}_{xxx}(\omega,\omega)$ and linear optical conductivity $\tilde{\sigma}^{\textrm{inter}}_{xx}(\omega)$, respectively. In Fig.~2, we find that if the chemical potential $\mu$ is smaller than half the gap $M$, the onset frequency is $M$. Because the chemical potential $\mu$ lies in the gap, the minimum energy for the inter-band transition is $2M$. The energy of absorbing two photons is $2\omega$, so that the onset frequency $2\omega\ge2M$. If the chemical potential $\mu$ is larger than $M$, the onset frequency is the chemical potential $\mu$, because in this case $2\omega\ge 2\mu$. There is a small drop in the nonlinear optical conductivity at $\omega=2\mu$, because this is the onset frequency for another inter-band transition involving one photon absorbing. Thus the number of photons in the process of frequency doubling decreases. For $\omega\geq2\mu$, curves with different values of $\mu$ fall on top of each other. 

In Fig.~3 we find that the onset frequency of the linear optical conductivity is $2\mu$, in contrast to the onset frequency $\mu$ of the nonlinear optical conductivity. When the frequency $\omega$ is larger than the onset frequency $2\mu$, the linear optical conductivity warps up, in contrast to the nonlinear optical conductivity which decreases as the frequency increases. Curves with different values of $\mu$ also fall on top of each other for the linear optical conductivity. In Fig.~4 and Fig.~5 we show the corresponding imaginary parts of the inter-band optical conductivity $\tilde{\sigma}_{xxx}^{\textrm{inter}}(\omega,\omega)$ and $\tilde{\sigma}_{xx}^{\textrm{inter}}(\omega)$: respectively.  The absolute value of the imaginary part of the nonlinear conductivity decreases to zero faster than that of the linear conductivity. 

\section{Conclusion}

In conclusion, we developed a method based on the trace of the velocity operator and Green's function to calculate the nonlinear response functions
in a non-centrosymmetric topological insulator. Our method is equivalent to the velocity matrix element method \cite{Bloemb} if a two-band free-electron approximation was considered. We obtained the nonlinear conductivity for frequency up-conversion in the second harmonic generation. In the model
used here (the two-band hexagonal warping model), the energy scale is
around 200 meV in the far-infrared region, relevant for the thermal energy. This model describes surface states of 3D TI. If the 3D TI $(\textrm{Bi}_{1-y}\textrm{Sb}_{y})_{2}\textrm{Te}_{3}$
(BST) is doped with magnetic impurities, a small gap is opened
in $\textrm{Cr}_{x}(\textrm{Bi}_{1-y}\textrm{Sb}_{y})_{2\text{\textminus}x}\textrm{Te}_{3}$
(CBST) \cite{Tokura,Yasuda} thus the inversion symmetry was broken. 

\begin{acknowledgments}
The authors thank Chong Wang, Yong Xu, A. W. Frisk Kockum, Mauro Cirio and Zubin Jacob
for useful discussions. Z. L. acknowledges the support of a JSPS Foreign
Postdoctoral Fellowship under Grant No. PE14052 and P16027. F.N. is supported in part by the: MURI Center for Dynamic Magneto-Optics via the
Air Force Office of Scientific Research (AFOSR) (FA9550-14-1-0040), Army Research Office (ARO) (Grant No. W911NF-18-1-0358),
Asian Office of Aerospace Research and Development (AOARD) (Grant No. FA2386-18-1-4045),
Japan Science and Technology Agency (JST) (the Q-LEAP program, the ImPACT program and CREST Grant No. JPMJCR1676),
Japan Society for the Promotion of Science (JSPS) (JSPS-RFBR Grant No. 17-52-50023, JSPS-FWO Grant No. VS. 059.18N),
RIKEN-AIST Challenge Research Fund, and the John Templeton Foundation.

\end{acknowledgments}

\appendix

\section{Derivation of the nonlinear optical conductivity from the Green's function technique }

In this appendix we present a general formula for the
calculation of the nonlinear conductivity tensor. We also expand the
imaginary frequency Green's function into a sum of the real frequency
spectral function which can be measured directly from ARPES experiments.
From this expansion we derive the intra-band and inter-band contribution
to the nonlinear conductivity. We present how to perform the sum in
imaginary frequency and obtain concise results in the non-interacting
electron approximation. Consider the hexagonal warping model as an example. The Green's function can be rewritten in the basis of Pauli matrices

\begin{equation}
\widehat{G}(\mathbf{k},i\omega_{n})=\frac{1}{2}\sum_{s=\pm}(1+s\mathbf{F}_{k}\cdot\mathbf{\sigma})G_{0}(\mathbf{k},s,i\omega_{n}),
\end{equation} 

where 
\begin{eqnarray*}
\mathbf{F}_{k}=\frac{(-v_{k}k\sin\theta,v_{k}k\cos\theta,\lambda k^{3}\cos(3\theta)+M)}{\sqrt{v_{k}^{2}k^{2}+[\lambda k^{3}\cos(3\theta)+M]^{2}}}\,,
\end{eqnarray*}
and $G_{0}(\mathbf{k},s,i\omega_{n})=[{i\omega_{n}+\mu-sE}]^{-1}$,
where $s=\pm1$ and the energy spectrum is given by 
\begin{equation*}
E=\sqrt{v_{k}^{2}k^{2}+[\lambda k^{3}\cos(3\theta)+M]^{2}}\,.
\end{equation*}

The velocity operator can be obtained as  (for simplicity we set $\hbar=1$ )
\begin{eqnarray}
v_{x} & = & \frac{\partial H_{0}}{\partial k_{x}}=v_{k}\sigma_{y}+\frac{\lambda}{2}(3k_{+}^{2}+3k_{-}^{2})\sigma_{z}\notag\\
 & = & v_{k}\sigma_{y}+3\lambda k^{2}\cos(2\theta)\sigma_{z}\notag\\
v_{y} & = & \frac{\partial H_{0}}{\partial k_{y}}=-v_{k}\sigma_{x}+\frac{\lambda}{2}(3ik_{+}^{2}-3ik_{-}^{2})\sigma_{z}\notag\\
 & = & -v_{k}\sigma_{x}-3\lambda k^{2}\sin(2\theta)\sigma_{z}\,.
\end{eqnarray}
In general $v_{x}=a_{0}+\mathbf{a}\cdot\mathbf{\sigma}$ , $v_{y}=b_{0}+\mathbf{b}\cdot\mathbf{\sigma}$,
and $\widehat{G}(\mathbf{k},i\omega_{n})=g_{0}+\mathbf{g}\cdot\mathbf{\sigma}$.
If we define $\mathbf{A}=(a_{0},\mathbf{a}),\mathbf{B}=(b_{0},\mathbf{b}),\mathbf{G}=(g_{0},\mathbf{g})$,
and use the following rules for dot and cross product of two vectors
\[
\mathbf{A\cdot B=}a_{0}b_{0}+\mathbf{a}\cdot\mathbf{b}\,,
\]
\[
\mathbf{A\times B=}a_{0}\mathbf{b}+b_{0}\mathbf{a}+i(\mathbf{a}\times\mathbf{b)}\,.
\]
Then the products of $v_{x}\widehat{G}(\mathbf{k,}\omega_{1})$ can
be evaluated as 
\begin{widetext}
\begin{align*}
 & (a_{0}+\mathbf{a}\cdot\mathbf{\sigma})(g_{01}+\mathbf{g_{1}}\cdot\mathbf{\sigma})=a_{0}g_{01}+g_{01}\mathbf{a}\cdot\mathbf{\sigma}+a_{0}\mathbf{g_{1}}\cdot\mathbf{\sigma}+\mathbf{a}\cdot\mathbf{g_{1}}+i(\mathbf{a}\times\mathbf{g_{1}})\cdot\mathbf{\sigma}\\
 & =\mathbf{A\cdot G_{1}}+\mathbf{(A\times G_{1})\cdot\mathbf{\sigma}}\,.
\end{align*}
The trace can be carried out in general as 
\begin{align*}
 & \mathrm{Tr}\langle(a_{0}+\mathbf{a}\cdot\mathbf{\sigma})(g_{01}+\mathbf{g_{1}}\cdot\mathbf{\sigma})(a_{0}+\mathbf{a}\cdot\mathbf{p)}(g_{02}+\mathbf{g}_{2}\cdot\mathbf{\sigma})(a_{0}+\mathbf{a}\cdot\mathbf{\sigma})(g_{03}+\mathbf{g_{3}}\cdot\mathbf{\sigma})\rangle\\
= & \mathrm{Tr}\langle(\mathbf{A\cdot G_{1}}+\mathbf{(A\times G_{1})\cdot\mathbf{\sigma}})(\mathbf{\mathbf{A\cdot G_{2}}+\mathbf{(A\times G_{2})\cdot\mathbf{\sigma}})}(\mathbf{A\cdot G_{3}}+\mathbf{(A\times G_{3})\cdot\mathbf{\sigma}})\rangle\\
= & [(\mathbf{A\cdot G_{1})}\mathbf{(A\cdot G_{2}})+(A\times G_{1})\cdot(A\times G_{2})](\mathbf{A\cdot G_{3}})+\mathbf{[(A\cdot G_{2}})\mathbf{(A\times G_{1})+(\mathbf{A\cdot G_{1}})\mathbf{(A\times G_{2})}]}\cdot(A\times G_{3})\\
 & +i(A\times G_{1})\times(A\times G_{2})\cdot(A\times G_{3})\,.
\end{align*}
The matrix Green's function $\widehat{G}(\mathbf{k},i\omega_{n})$
can be conveniently written in terms of a matrix spectral function
$\widehat{A}(\mathbf{k},\omega)$ with 
\begin{equation}
\widehat{G}(\mathbf{k},i\omega_{n})=\int_{-\infty}^{\infty}\frac{d\omega}{2\pi}\frac{\widehat{A}(\mathbf{k},\omega)}{i\omega_{n}-\omega}\,,
\end{equation}
then the conductivity in the long-wavelength limit becomes 
\begin{eqnarray}
\tilde{\sigma}_{xxx}(\omega,\omega) & = & \frac{e^{3}}{\omega^{2}}\frac{i}{4\pi^{2}}\int_{0}^{k_{\textrm{cut}}}kdkd\theta\int_{-\infty}^{\infty}\frac{d\omega_{1}}{2\pi}\int_{-\infty}^{\infty}\frac{d\omega_{2}}{2\pi}\int_{-\infty}^{\infty}\frac{d\omega_{3}}{2\pi}\nonumber \\
 & \times & T\sum_{l}\frac{1}{i\omega_{l}-\omega_{1}}\frac{1}{i\omega_{l}+i\omega_{n}-\omega_{2}}\frac{1}{i\omega_{l}-i\omega_{n}-\omega_{3}}\notag\\
 & \times & \mathrm{Tr}\langle v_{x}\widehat{A}(\mathbf{k,}\omega_{1})v_{x}\widehat{A}(\mathbf{k,}\omega_{2})v_{x}\widehat{A}(\mathbf{k,}\omega_{3})\rangle_{i\omega_{n}\rightarrow\omega+i\delta}\,,
\end{eqnarray}

For two-band models, the spectral function $\widehat{A}(\mathbf{k},\omega)$
can be expanded in the basis of Pauli matrices, 
\begin{equation*}
\widehat{A}(\mathbf{k},\omega)=A_{I}(\mathbf{k,}\omega)+A_{x}(\mathbf{k,}\omega)\sigma_{x}+A_{y}(\mathbf{k,}\omega)\sigma_{y}+A_{z}(\mathbf{k,}\omega)\sigma_{z}\,.
\end{equation*}
In the free-electron approximation (ignoring impurity scattering and
electron-phonon scattering), the spectral functions are given by 
\begin{equation}
A_{I}(\mathbf{k,}\omega)=\delta(\omega+\mu-E)+\delta(\omega+\mu+E)\,,
\end{equation}
\begin{equation}
A_{x}(\mathbf{k,}\omega)=\frac{-v_{k}k\sin\theta[\delta(\omega+\mu-E)-\delta(\omega+\mu+E)]}{\sqrt{v_{k}^{2}k^{2}+[\lambda k^{3}\cos(3\theta)+M]^{2}}}\,,
\end{equation}
\begin{equation}
A_{y}(\mathbf{k,}\omega)=\frac{v_{k}k\cos\theta[\delta(\omega+\mu-E)-\delta(\omega+\mu+E)]}{\sqrt{v_{k}^{2}k^{2}+[\lambda k^{3}\cos(3\theta)+M]^{2}}}\,,
\end{equation}
\begin{equation}
A_{z}(\mathbf{k,}\omega)=\frac{[\lambda k^{3}\cos(3\theta)+M][\delta(\omega+\mu-E)-\delta(\omega+\mu+E)]}{\sqrt{v_{k}^{2}k^{2}+[\lambda k^{3}\cos(3\theta)+M]^{2}}}\,.
\end{equation}
The trace can be carried out as 
\begin{align}
 & \mathrm{Tr}\langle v_{x}\widehat{A}(\mathbf{k,}\omega_{1})v_{x}\widehat{A}(\mathbf{k,}\omega_{2})v_{x}\widehat{A}(\mathbf{k,}\omega_{3})\rangle \notag\\
= & 8F(k,\theta)[\delta(\omega_{1}+\mu-E)\delta(\omega_{2}+\mu+E)\delta(\omega_{3}+\mu-E)+\delta(\omega_{1}+\mu+E)\delta(\omega_{2}+\mu-E)\delta(\omega_{3}+\mu-E)\notag\\
 & +\delta(\omega_{1}+\mu-E)\delta(\omega_{2}+\mu-E)\delta(\omega_{3}+\mu+E)-\delta(\omega_{1}+\mu-E)\delta(\omega_{2}+\mu+E)\delta(\omega_{3}+\mu+E)\notag\\
 & -\delta(\omega_{1}+\mu+E)\delta(\omega_{2}+\mu-E)\delta(\omega_{3}+\mu+E)-\delta(\omega_{1}+\mu+E)\delta(\omega_{2}+\mu+E)\delta(\omega_{3}+\mu-E)]\notag\\
+ & 8F_{\textrm{intra}}(k,\theta)[\delta(\omega_{1}+\mu-E)\delta(\omega_{2}+\mu-E)\delta(\omega_{3}+\mu-E)-\delta(\omega_{1}+\mu+E)\delta(\omega_{2}+\mu+E)\delta(\omega_{3}+\mu+E)]\,,
\end{align}
where we have defined two functions, 
\begin{eqnarray}
F(k,\theta) & = & \frac{[k_{x}v_{k}^{2}+3\lambda(k_{x}^{2}-k_{y}^{2})(M+\lambda k_{x}(k_{x}^{2}-3k_{y}^{2}))]}{[v_{k}^{2}k^{2}+(\lambda k^{3}\cos(3\theta)+M)^{2}]^{3/2}}\notag\\
 & \times & v_{k}^{2}[v_{k}^{2}k_{y}^{2}+\lambda^{2}(4k_{x}^{6}+9k_{x}^{4}k_{y}^{2}-18k_{x}^{2}k_{y}^{4}+9k_{y}^{6})\notag\\
 &  & -4\lambda Mk_{x}^{3}+M^{2}]\,,
\end{eqnarray}
\begin{eqnarray}
F_{\textrm{intra}}(k,\theta) & = & \frac{[k_{x}v_{k}^{2}+3\lambda(k_{x}^{2}-k_{y}^{2})(M+\lambda k_{x}(k_{x}^{2}-3k_{y}^{2}))]}{[v_{k}^{2}k^{2}+(\lambda k^{3}\cos(3\theta)+M)^{2}]^{3/2}}\notag\\
 & \times & [k_{x}(v_{k}^{2}+3\lambda^{2}(k_{x}^{4}-4k_{x}^{2}k_{y}^{2}+3k_{y}^{4}))+3\lambda M(k_{x}^{2}-k_{y}^{2})]^{2}\,.
\end{eqnarray}
These terms can be separated into inter-band and intra-band contributions
to the nonlinear conductivity. 

\section{Intra-band nonlinear conductivity}

The intra-band nonlinear conductivity
includes those terms proportional to $\delta(\omega_{1}+\mu-E)\delta(\omega_{2}+\mu-E)\delta(\omega_{3}+\mu-E)$
and $\delta(\omega_{1}+\mu+E)\delta(\omega_{2}+\mu+E)\delta(\omega_{3}+\mu+E)$,
which will contribute to the zero-frequency DC conductivity. Performing
the sum over Matsubara frequencies, we obtain 
\begin{align*}
 & T\sum_{l}\frac{1}{i\omega_{l}-\omega_{1}}\frac{1}{i\omega_{l}+i\omega_{n1}-\omega_{2}}\frac{1}{i\omega_{l}-i\omega_{n2}-\omega_{3}}\\
 & =T\sum_{l}\frac{1}{i\omega_{n1}-\omega_{2}+\omega_{1}}\Big(\frac{1}{i\omega_{l}-\omega_{1}}-\frac{1}{i\omega_{l}+i\omega_{n1}-\omega_{2}}\Big)\frac{1}{i\omega_{l}-i\omega_{n2}-\omega_{3}}\\
 & =T\sum_{l}\frac{1}{i\omega_{n1}-\omega_{2}+\omega_{1}}\Big[\frac{1}{-i\omega_{n2}-\omega_{3}+\omega_{1}}\Big(\frac{1}{i\omega_{l}-\omega_{1}}-\frac{1}{i\omega_{l}-i\omega_{n2}-\omega_{3}}\Big)\\
 & -\frac{1}{-i\omega_{n1}-i\omega_{n2}-\omega_{3}+\omega_{2}}\Big(\frac{1}{i\omega_{l}+i\omega_{n1}-\omega_{2}}-\frac{1}{i\omega_{l}-i\omega_{n2}-\omega_{3}}\Big)\Big]\\
 & =\frac{1}{i\omega_{n1}-\omega_{2}+\omega_{1}}\Big[\frac{f(\omega_{3})-f(\omega_{1})}{i\omega_{n2}+\omega_{3}-\omega_{1}}+\frac{f(\omega_{2})-f(\omega_{3})}{i\omega_{n1}+i\omega_{n2}+\omega_{3}-\omega_{2}}\Big]\,.
\end{align*}
And the intra-band conductivity becomes 
\begin{eqnarray}
\tilde{\sigma}_{xxx}(\omega,\omega)_{\textrm{intra}} & = & \frac{e^{3}}{\omega^{2}}\frac{8i}{4\pi^{2}}\int_{0}^{k_{\textrm{cut}}}kdkd\theta\int_{-\infty}^{\infty}\frac{d\omega_{1}}{2\pi}\int_{-\infty}^{\infty}\frac{d\omega_{2}}{2\pi}\int_{-\infty}^{\infty}\frac{d\omega_{3}}{2\pi}F_{\textrm{intra}}(k,\theta)\notag\\
 & \times & [\delta(\omega_{1}+\mu-E)\delta(\omega_{2}+\mu-E)\delta(\omega_{3}+\mu-E)-\delta(\omega_{1}+\mu+E)\delta(\omega_{2}+\mu+E)\delta(\omega_{3}+\mu+E)]
\notag\\
 & \times &  \frac{1}{\omega+i\delta-\omega_{2}+\omega_{1}}\Big[\frac{f(\omega_{3})-f(\omega_{1})}{\omega+i\delta+\omega_{3}-\omega_{1}}+\frac{f(\omega_{2})-f(\omega_{3})}{2\omega+2i\delta+\omega_{3}-\omega_{2}}\Big]\,,
 \end{eqnarray}
the intra-band conductivity can be numerically evaluated, by replacing
the $\delta$ function with the broadened Lorentzian function. One
can also evaluate the intra-band conductivity analytically; one example
was given in the appendix of Ref.~[\onlinecite{Li1}]. 

\section{Inter-band nonlinear conductivity}
The other terms like $\delta(\omega_{1}+\mu-E)\delta(\omega_{2}+\mu+E)\delta(\omega_{3}+\mu-E)$
are included in the inter-band nonlinear conductivity which will contribute
to the nonzero-frequency AC conductivity, written as 

\begin{eqnarray*}
\tilde{\sigma}_{xxx}(\omega,\omega)_{\textrm{inter}} & = & \frac{e^{3}}{\omega^{2}}\frac{8i}{4\pi^{2}}\int_{0}^{k_{\textrm{cut}}}kdkd\theta F(k,\theta)\Big[\frac{1}{\omega+i\delta+2E}\Big(\frac{f(-E)-f(E)}{2\omega+2i\delta+2E}\Big)+\frac{1}{\omega+i\delta-2E}\Big(\frac{f(E)-f(-E)}{\omega+i\delta+2E}\Big)\notag\\
 & + & \frac{1}{\omega+i\delta}\Big(\frac{f(-E)-f(E)}{\omega+i\delta-2E}+\frac{f(E)-f(-E)}{2\omega+2i\delta-2E}\Big)-\frac{1}{\omega+i\delta+2E}\Big(\frac{f(-E)-f(E)}{\omega+i\delta-2E}\Big)\notag\\
 & - & \frac{1}{\omega+i\delta-2E}\Big(\frac{f(E)-f(-E)}{2\omega+2i\delta-2E}\Big)-\frac{1}{\omega+i\delta}\Big(\frac{f(E)-f(-E)}{\omega+i\delta+2E}+\frac{f(-E)-f(E)}{2\omega+2i\delta+2E}\Big)\Big]\,,
\end{eqnarray*}
which is further simplified as
\begin{eqnarray*}
\tilde{\sigma}_{xxx}(\omega,\omega)_{\textrm{inter}} & = & \frac{2ie^{3}}{\omega^{2}\pi^{2}}\int_{0}^{k_{\textrm{cut}}}kdkd\theta F(k,\theta)\notag\\
 & \times & \Big[\frac{2}{\omega+i\delta-2E}\frac{f(E)-f(-E)}{\omega+i\delta+2E}-\frac{f(E)-f(-E)}{\omega+i\delta-2E}\frac{1}{\omega+i\delta-E}-\frac{f(E)-f(-E)}{\omega+i\delta+2E}\frac{1}{\omega+i\delta+E}\Big]\,.
\end{eqnarray*}
Finally we obtained 
\begin{eqnarray}
\tilde{\sigma}_{xxx}(\omega,\omega)_{\textrm{inter}} & = & \frac{ie^{3}}{\omega^{2}\pi^{2}}\int_{0}^{2\pi}d\theta\int_{0}^{k_{\textrm{cut}}}kdk\frac{[f(E)-f(-E)]}{E}\notag\\
 & \times F(k,\theta) & \Big[\frac{1}{\omega+i\delta+2E}-\frac{1}{\omega+i\delta-2E}+\frac{2}{\omega+i\delta-E}-\frac{2}{\omega+i\delta+E}\Big]\,.
\end{eqnarray}

\end{widetext}

\end{document}